\documentclass{SCGE}
\usepackage{amssymb}
\usepackage{stmaryrd}
\usepackage{graphicx}
\usepackage{rotating}
\begin{document}

\begin{picture}(0,0){\rm
\put(0,-20){\makebox[160truemm][l]{\bf {\sanhao\raisebox{2pt}{.}}
News and Views  {\sanhao\raisebox{1.5pt}{.}}}}}
\end{picture}

\def\bm{\boldsymbol}

\def\dl{\displaystyle}
\def\du{\end{document}}
\def\d{{\rm d}}
\def\e{{\rm e}}
\def\i{{\rm i}}

\def\pi{{\uppi}}

\Year{2020} %
\Month{March} %
\Vol{63} 
\No{xx} 
\BeginPage{1} 
\AuthorMark{{\rm Xu}}  
\AuthorMarkCite{{\rm Xu}. } 
\DOI{https://doi.org/xxx} 
\ArtNo{xxx}

\title[3-f in triangle]{Light-quark flavours in a triangle}

\author[]{Renxin XU}{}

\address[]{School of Physics and Kavli Institute for Astronomy and Astrophysics, Peking University, Beijing 100871, China; r.x.xu@pku.edu.cn}

\maketitle \vspace{-3.5mm}{\footnotesize\begin{center}
Received xxx, 2020; accepted xxx; published online xxx
\end{center}}\vspace*{-5mm}



\begin{center}
%
{\color{blue} \em All of the light flavors seem to work well for bulk strong matter at pressure free.}\\
\Cit{R. X. Xu, Light-quark flavours in a triangle, Sci. China-Phys. Mech. Astron. {\bf xx}, xxx (2020), https://doi.org/xxx}%
\end{center}

\textwidth=178truemm \textheight=236truemm

\wuhao\vspace*{1.5mm}

\begin{multicols}{2}

\renewcommand{\baselinestretch}{1.08} \baselineskip 12.2pt\parindent=10.8pt

\renewcommand{\thefootnote}


\noindent
The coupling between {\em material units}, either weak or strong,
determines the nature of matter controlled by different forces.
As for three fundamental interactions (gravitational,
electromagnetic, and strong), all the gauge bosons to mediate the
forces are massless, and we may simply approximate the interaction
potential energy at short distances with a Coulomb-like form of
\begin{equation}
V = -\alpha_i/r,
\label{Coulomb}%
\end{equation}
where $r$ is the separation between material units, and $\alpha_i$ measures the coupling strength of the fundamental forces, with $i=\{{\rm g, e, s}\}$ for gravitational, electromagnetic and strong interactions, respectively.

For the {\em gravity}, the mass of a building unit, $m$, should be
large, otherwise the interaction would be negligible.
In this sense, the angular momentum of a unit inside a
gravity-controlled system (e.g., the solar system) is always much
larger than the Planck constant, $\hbar$, and it is then safe that
we can neglect totally the quantum effect (i.e., $\hbar \sim 0$) in
celestial mechanics (including galactic dynamics).
Meanwhile, massive units inside this system should usually move
non-relativistically, except for compact binary merger.

For the {\em electromagnetism}, however, quantum effects can not be  negligible anymore, because the angular momentum could be so small that it is comparable to Planck-$\hbar$, in view of a higher $\alpha_{\rm e}$ and thus of a smaller length $\ell_{\rm sys}$.
According to the Heidelberg relation, $p_{\rm ground}\cdot \ell_{\rm
sys}\sim \hbar$, where $p_{\rm ground}$ is the momentum of a unit at a ground
state, and its kinematic energy is then $E_{\rm k}\simeq
\hbar^2/(m\ell_{\rm sys}^2)$ if the unit moves also
non-relativistically.
Approximating the binding-energy, $\alpha_i/\ell_{\rm sys}$, to the
kinematic energy $E_{\rm k}$, one has $\ell_{\rm sys}\simeq
\hbar^2/(\alpha_i m)\propto \alpha_i^{-1}$.
This tells us that gravity-controlled objects (e.g., galaxies) should be much larger than typical electric (condensed) matter.
For the simplest electromagnetism-bound system of a real hydrogen
(electron mass $m_{\rm e}$, proton mass $m_{\rm p}$) with a realistic interaction energy of $E_{\rm
ep}\simeq -e^2/\ell_{\rm sys}$, one has
$%
\ell_{\rm ep} \sim {\hbar^2/(m_{\rm e}e^2)} = \alpha_{\rm em}^{-1}
\cdot \hbar c/(m_{\rm e}c^2),
$%
where $\alpha_{\rm em}=e^2/(\hbar c)\simeq 1/137$ is the coupling
constant of the electromagnetic interaction.
Therefore, we could estimate the typical density of electric
matter (i.e., atom matter) as~\cite{Xu2018}
\begin{equation}
\rho_{\rm EM}\simeq {m_{\rm p}\over \ell_{\rm ep}^3} =
({\alpha_{\rm em}c \over \hbar})^3 m_{\rm e}^3m_{\rm p} \sim 10~{\rm g/cm^3}.
\label{rho_EM}%
\end{equation}

For the {\em strong} interaction, in addition to the quantum effects, particle-antiparticle pairs would also play an essential role for a strong unit (i.e., hadron), in view of a much higher
$\alpha_{\rm s}$ and thus of an extremely small $\ell_{\rm sys}$.
In sharp contrast to the case of atoms, because of small $\ell_{\rm sys}$ and thus high momentum, quarks would be relativistic (even extremely) so that gluons, coupled with quarks, can split into a number of quark-antiquark pairs, named sea quarks.
Therefore, it is then recognized that a baryon contains 3 valence
quarks, sea quarks as well as gluons.
It is worth noting that the strange flavour of sea quarks ($s{\bar
s}$) play also a fundamental role in the structure of nucleon
though proton and neutron are certainly
non-strange~\cite{Ellis2001}.
This is simply due to the fact that the energy scale of strong
matter at pressure free is of order~\cite{Xu2018} $E_{\rm scale}\sim 0.5$ GeV $>
m_{\rm s}>m_{\rm u,d}$, where
$m_{\rm s}$ is the current mass of strange quark and $m_{\rm u,d}$ of
up or down quarks.
A perturbative calculation of quantum chromodynamics (QCD) may
predict a nucleon sea with light-flavour symmetry, but the observed flavour asymmetry
in the light-quark sea would be the result of the
non-perturbative nature.
Nevertheless, an effective model is applied to describe hadrons, in
which constituent-quark degrees of freedom interacting via a
non-relativistic potential are assumed, though the constituent quark
masses ($\tilde{m}_{\rm q}$ for a q-flavour quark) are much larger than their real (current) masses.
For non-relativistic constituent quarks interacting with others via the Coulomb-like
form of Eq.(\ref{Coulomb}), in an analogy of
Eq.(\ref{rho_EM}), the density of strong matter,
$%
\rho_{\rm SM}\simeq(\alpha_{\rm c}c/\hbar)^3 \tilde{m}_{\rm q}^4 \sim 2\times 10^{15}~{\rm g/cm^3}
$, %
could be representative of the nuclear density $\rho_0$, where
$\tilde{m}_{\rm q}=0.3$ GeV and the coupling constant of the strong
interaction $\alpha_{\rm c}=1$ are chosen.

We would emphasize that, because of the asymmetry of
$e^\pm$, virtual {\em strange} quarks in the nucleon
sea could materialize as valence ones when normal baryonic
matter in the core of an evolved massive star is squeezed so great
that nuclei come in close contact.
For lepton-related weak interactions of $u+e^-\rightarrow s+\nu_e$ and ${\bar u}+e^+\rightarrow {\bar s}+{\bar \nu}_e$, the former should be more effective than the latter to to kill
electrons more and more energetic in the collapsing core, producing eventually valence strange
quarks as many as the up and down quarks.$^*$\footnote{%
$^*$
This is for light flavours only, not applicable for heavy ones ($c,t,b$).
In order to effectively cancel energetic electrons, we need two flavours charged negatively (-1/3) when introducing one flavour charged positively (+2/3), if we try to reach approximately a flavour balance.
} %
An alternative way is for two flavours only ($u+e^-\rightarrow d+\nu_e$ and ${\bar u}+e^+\rightarrow {\bar d}+{\bar \nu}_e$), resulting in an extremely asymmetric state of isospin (Landau's neutron-rich `gigantic nucleus' as an example).
In the light that Nature might love symmetry, we would focus on three flavours ($u,d,s$) in this essay, investigating in a triangle diagram explained as following.

\begin{figure}[H]
\begin{center}
\includegraphics[width=0.3\textwidth]{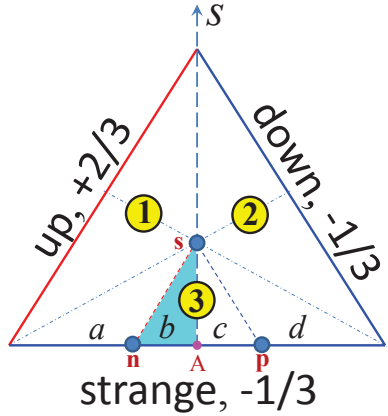}
\hfill
\end{center}
\caption{%
Triangle of light-quark flavours. A point inside the triangle defines a state with certain quark numbers of three flavours (\{$n_{\rm u}, n_{\rm d}, n_{\rm s}$\} for up, down and strange quarks), which are measured by the heights of the point to one of the triangle edges.
Point ``s'' is the center of the triangle, at which one has $n_{\rm u}=n_{\rm d}=n_{\rm s}$.
Line ``sn'' is parallel to the up edge, while line ``sp'' to the down edge.
Axis $S$ is for strangeness, where the isospin symmetry is also perfect.
The triangle is divided into three regions (\textcircled{1}, \textcircled{2}, \textcircled{3}), and furthermore the region \textcircled{3} into $a,b,c$ and $d$.
}%
\label{triangle}
\end{figure}
{\em The triangle}.
Due to baryon conservation, it is convenient to discuss the quark numbers of three flavours in a regular triangle (Fig.~\ref{triangle}), for given baryon density, $n_{\rm b}=(n_{\rm u}+n_{\rm d}+n_{\rm s})/3$.
It is evident that the bottom strange edge is divided into three equal parts
by points ``n'' and ``p''  because the triangle ``$\triangle$snp''
is left-right symmetrical to the ``$S$''-axis but shrinks by
two-thirds.
Those states in a line parallel to that of ``ns'' have a constant
number density of up quark, and thus a constant charge density.
The up-quark accounts for one-third at the
states in the line of ``ns'' (2$n_{\rm u}=n_{\rm d}+n_{\rm s}$), so that no electrons are necessary to
keep neutrality.
However, the electron population increases at states in the right of
line ``ns'', until to the maximum at the state point ``p'' (a state
in the right of ``p'' is unrealistic, see below).
For the sake of simplicity, the triangle is divided into three
regions: \textcircled{1} ($n_{\rm u}<n_{\rm d}$ and $n_{\rm
u}<n_{\rm s}$), \textcircled{2} ($n_{\rm d}<n_{\rm u}$ and $n_{\rm
d}<n_{\rm s}$) and \textcircled{3} ($n_{\rm s}<n_{\rm u}$ and
$n_{\rm s}<n_{\rm d}$).

Normal nuclei in vacuum are non-strange, with almost equal numbers
of up and down quarks due to the nuclear symmetry energy, so that they
sit around point ``A''.
However, as the electron density in the environment increases (e.g.,
by squeezing normal matter), strangeness (valence $s$-quark) becomes
more and more significant, but its density $n_{\rm s}$ should be the
smallest (i.e., $n_{\rm s}<n_{\rm u}$ and $n_{\rm s}<n_{\rm d}$ in
the region \textcircled{3}).
We note that point
``n'' for $2n_{\rm u} = n_{\rm d}$ but $n_{\rm s}=0$ (e.g., pure
neutron matter), and that point ``s'' for $n_{\rm u} = n_{\rm d}=n_{\rm
s}$ (e.g., strange quark matter with 3-flavour symmetry).
The total charge of quarks is negative in region ``a'', but positive in
regions ``b'', ``c'' and ``d''.
If Nature loves the isospin symmetry, though with symmetry broken slightly, both regions of ``a'' and ``d'' could be unlikely because of $n_{\rm u}<n_{\rm b}<n_{\rm d}$ in ``a'' and of  $n_{\rm d}<n_{\rm b}<n_{\rm u}$ in ``d'', and a realistic state could thus be in either ``b'' and ``c''.
This means that the most positively charged state of quarks (i.e., the
electron density $n_{\rm e}$ to be the highest) would then be at point
``p'': $n_{\rm d}=n_{\rm u}/2=n_{\rm b}$ but $n_{\rm s}=0$ (a pure ``proton''
star if only nucleon degree of freedom keeps).

\begin{figure}[H]
\begin{center}
\includegraphics[width=0.46\textwidth]{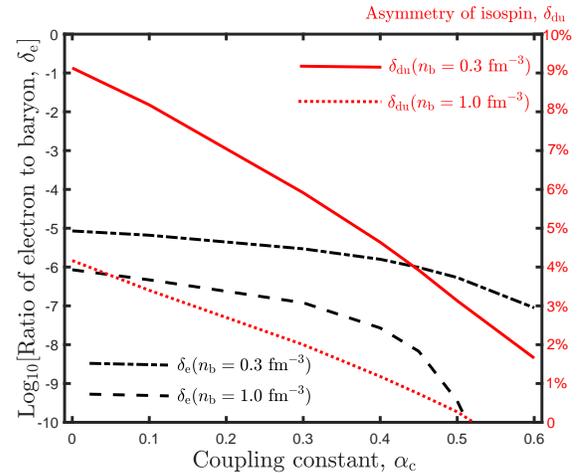}
\hfill
\end{center}
\caption{%
Light-quark flavours become more symmetric as the strong coupling, $\alpha_{\rm c}$, increases.
Y-axis in the left shows the ratio of electron to baryon number density, $\delta_{\rm e}=n_{\rm e}/n_{\rm b}$, while that in the right indicates the isospin asymmetry, $\delta_{\rm du}=(n_{\rm
d}-n_{\rm u})/n_{\rm b}$.
It is evident that the 3-flavour symmetry recovers more perfectly at higher coupling.
}%
\label{bag}
\end{figure}
Where are stable states of strong matter in the triangle?
We know certainly that a normal nucleus take a seat around point ``A'',
but what if normal baryonic matter is compressed to supranuclear
density?
There are many speculations related, including nucleon star (i.e., ‘gigantic nucleus’ initialized by Lev Landau), hyperon star, strange quark star, etc.
Two essential issues are frequently focused in the study: quarks deconfined or not? strangeness significant or not?

For infinite strong matter with negligible surface effect, a
bag-linked model, or the bag crystal model (e.g.,~\cite{Zhang1992}),
would be adaptable for both cases of two and three quark-flavors,
since the dynamics of quarks not localized in separated bags would introduce
effective interactions so as to model condensed strong-matter.
Surely it is a matter of non-perturbative QCD to know the exact
state of strong matter at different density.
Nevertheless, the bag-linked model could help for an
order-of-magnitude estimation.
In the regime of 3-flavours of quarks, we simply apply the
thermodynamic potentials of conventional bag model (e.g, ~\cite{afo1986}) for both strange
quark matter (quarks free, e.g.~\cite{Witten1984}) and strangeon
matter (quarks localized almost in bag-like
strangeon,~\cite{Xu2003}), paying small attention to the difference of
bag boundary condition.
The calculated results are presented in Fig.~\ref{bag}, showing the
ratio of electron to baryon numbers ($\delta_{\rm e}=n_{\rm e}/n_{\rm b}$, to be representative of
3-flavour asymmetry) and the isospin asymmetry ($\delta_{\rm du}=(n_{\rm
d}-n_{\rm u})/n_{\rm b}$) as a function of the strong coupling
constant, $\alpha_{\rm c}$.
For gravity-free strong matter with a small baryon number (e.g., $n_{\rm
b}\simeq 0.3$/fm$^3$), the ratio of electron to baryon density, $\delta_{\rm e}$, could
be from $10^{-7}$ to $10^{-5}$.
Also, it is evident that the number of down-quark is more than that
of up-quark, at a level of 1\%$\sim$ 9\% for $n_{\rm b}=0.3$/fm$^3$.
Therefore, bulk strong matter could be in region `b' of
Fig.~\ref{triangle}, closing to the point ``s''.
Stronger coupling (i.e., larger $\alpha_{\rm c}$) would result in a more symmetric state with 3-flavours of quarks.

{\em Cosmic manifestations of bulk strong matter}.
Given the difficulty and uncertainty of QCD's calculations including rich nonperturbative  effects, it is then indispensable to trace evidence from cosmic manifestations.
Compact stars provides a testbed. They should not be at point ``p'' (proton star) due to energetic electrons with kenematic energy of $\sim 10^2$ MeV, but could be either around point ``n'' (neutron star) or point ``s'' (strange star).
In addition to various astrophysical arguments (e.g., maximum-mass dependent state), it is still worth noting that protons would be extracted from the surface of neutron star but not from (bare) strange star, resulting in possible observational differences of magnetospheric activity.
Heavy nuclei (e.g., iron) in active polar cap could be disintegrated into nucleons by the bombardment of energetic e$^\pm$ pairs ($>$ TeV), and this electron-disintegration reaction may not favor a bare neutron star surface even with strong magnetic field.
We may expect Chinese FAST, the biggest single-dish radio telescope~\cite{Jiang2019}, to identify a clear signature revealing the differences (e.g., via single pulses~\cite{Lu2019}).

Strong nuggets passing through the Earth? They could probably created either during cosmic separation of QCD phases (as dark matter candidate) or while a strange star forms (via supernova or merger), providing an alternative test.
Certainly, neutron nugget (the asymmetric state at point ``n'') cannot exist at pressure free, but strangeon nugget can.
We point out that a strangeon nugget would be magnetized because of ferromagnetism of electrons~\cite{LX2016}, resulting in a larger cross-section of reaction between the nugget and normal baryon matter.
Only a tiny fraction of electrons, order of $3(3\pi^2)^{-1/3}\alpha_{\rm
em}\simeq 0.7\%$, contribute to the spin-alignment (a symmetric spin state, with an antisymmetric spatial wave function) so that the system becomes more stable due to a lower energy of the Coulomb interaction between electrons~\cite{LX2016}, and the magnetic moment of nugget with baryon number $N_{\rm b}$ could be $\mu_{\rm nugget}\sim 7\times 10^{-3}\mu_{\rm B}\delta_{\rm e}N_{\rm b}$ for a single magnetic domain, where $\mu_{\rm B}=9\times 10^{-21}$ Gauss$\cdot$cm$^3$ is the Bohr magneton of electron.
We can then estimate the magnetic field on strangeon nugget surface,
$B\sim 7\times 10^{-3}\mu_{\rm B}\delta_{\rm e}n_{\rm b}\sim 2\times
10^{11}$ G, for $\delta_{\rm e}=10^{-5}$ and $n_{\rm b}= 2\rho_0\simeq 0.3$ fm$^{-3}$
(see Fig.~\ref{bag}).
This field would take dynamical action inside an atom when $B>B_0=\alpha_{\rm em}^2B_{\rm Q}=2.4\times 10^9$ G, where $B_{\rm Q}=4.4\times 10^{13}$ G is the critical QED field strength, and the cross section of strangeon nugget is then order of $\sigma\sim 2\times 10^{-25}N_{\rm b}^{2/3}$ cm$^2$.
A factor of $1/\sqrt{\cal N}$ should be corrected for $\mu_{\rm nugget}$, if the number of magnetic domains inside strangeon matter is ${\cal N}$.

What if a strangeon nugget passes through the Earth?
For a strangeon nugget traveling a distance $L$ in medium with mass density $\rho_{\rm m}$, its momentum would be lost notably when $\sigma\cdot L\cdot \rho_{\rm m}\simeq N_{\rm b}m_{\rm p}$, i.e.,
\begin{equation}
L\simeq 8\times 10^4(N_{\rm b}/10^{12})^{1/3}{\rm cm}\cdot (1~{\rm g~cm^{-3}}/\rho_{\rm m}).
\label{L}%
\end{equation}
It is evident that strangeon nugget may slow down significantly in solid earth if $N_{\rm b}<10^{24}$, but could easily go across the atmosphere as $N_{\rm b}>10^9$ for stable nugget.
Both electromagnetic and hadronic cascades occur in an air shower for relativistic strangeon
nuggets, the energy loss of which determines whether a small nugget could stop inside the Earth.
A highly charged nugget might be accelerated effectively near an active pulsar, rather than by Galactic shock fronts.

{\em A summary}.
Bulk strong matter could be 3-flavoured in the triangle shown in
Fig.~\ref{triangle}, to be manifested diversely in the form of
compact star and strange nugget.

\vspace*{2mm} \noindent {\small \em This work is supported by the
National Key R\&D Program of China (No. 2017YFA0402602), the
National Natural Science Foundation of China (Grant Nos. 11673002
and U1531243), and the Strategic Priority Research Program of CAS
(No. XDB23010200).}

\end{multicols}


\begin{thebibliography}{99}

\bibitem{Xu2018}
R. Xu, Sci. China-Phys. Mech. Astron. 61, 109531 (2018)
arXiv:1802.04465.

\bibitem{Ellis2001}
J. R. Ellis, Nucl. Phys. A684, 53 (2001) arXiv:hep-ph/0005322.

\bibitem{Zhang1992}
Q. R. Zhang, H. M. Liu, Phys. Rev. C {\bf 46}, 2249 (1992).

\bibitem{afo1986}
C. Alcock, E. Farhi, and A. Olinto, ApJ. 310, 261 (1986).

\bibitem{Witten1984}
E. Witten, Phys. Rev. D {\bf 30}, 272 (1984).

\bibitem{Xu2003}
R. X. Xu, ApJ. {\bf 596}, L59 (2003) arXiv:astro-ph/0302165.

\bibitem{Jiang2019}
P. Jiang, Y. L. Yue, H. Q. Gan, et al., Sci. China-Phys. Mech. Astron. 62, 959502 (2019)  arXiv:1903.06324.

\bibitem{Lu2019}
J. G. Lu, B. Peng, R. X. Xu, et al., Sci. China-Phys. Mech. Astron. 62, 959505 (2019) arXiv:1903.06362.

\bibitem{LX2016}
X. Y. Lai, R. X. Xu, Chin. Phys. C 9, 095102 (2016) arXiv:1410.4949.

\end{thebibliography}
\end{document}